\documentclass{emulateapj}

\submitted{ }

\newcommand{\atan}{\mbox{arctan}}
\newcommand{\degK}{~\mbox{K}}
\newcommand{\grad}{^{\mbox{\small o}}}

\newcommand{\dyn}{\mbox{~dyn}}
\newcommand{\cm}{\mbox{~cm}}
\newcommand{\kpc}{\mbox{~kpc}}
\newcommand{\pc}{\mbox{~pc}}
\newcommand{\Myr}{\mbox{~Myr}}
\newcommand{\kms}{\mbox{~km s}^{-1}}
\newcommand{\microG}{\mu \mbox{G}}

\begin{document}

\title{3D MHD Modeling of the Gaseous Structure of the Galaxy:
  Description of the Simulations.}

\author{Gilberto C. G\'omez\altaffilmark{1}}
\affil{Department of Astronomy, University of Wisconsin-Madison,
  475 N. Charter St., Madison, WI 53706 USA}
\email{gomez@wisp.physics.wisc.edu}
\and
\author{Donald P. Cox}
\affil{Department of Physics, University of Wisconsin-Madison,
  1150 University Ave., Madison, WI 53706 USA}
\email{cox@wisp.physics.wisc.edu}

\altaffiltext{1}{Now at Department of Astronomy - University of Maryland,
  College Park, MD 20742; e-mail: gomez@astro.umd.edu}

\begin{abstract}

As discussed in \citet{gom02}, the extra stiffness that the magnetic
field adds to the ISM changes the way it reacts to the presence of a
spiral perturbation.  At intermediate to high $z$, the gas shoots up
before the arm, flows over, and falls behind it, as it approaches
the next arm.  This generates a multicell circulation pattern,
within each of which the net radial mass flux is positive near the
midplane and negative at higher $z$.  The flow distorts the magnetic
field lines.  In the arm region, the gas flows nearly parallel to
the arm, and therefore, the magnetic field adopts a similar pitch
angle.  Between the arms, the gas flows out in radius, generating a
negative pitch angle in the magnetic field.  The intensity and
direction of the field yield synthetic synchrotron maps that
reproduce some features of the synchrotron maps of external
galaxies, like the islands of emission and the displacement between
the gaseous and synchrotron arms.  When comparing the magnitude of
the field with the local gas density, two distinctive relations
appear, depending on whether the magnetic pressure is dominant.

Above the plane, the density structure develops a shape resembling a
breaking wave.  This structure collapses and rises again with a
period of about $60 \Myr$, similar to that of a vertical oscillation
mode.  The falling gas plays an important part in the overall
hydrostatics, since its deceleration compresses the low $z$ gas,
raising the average midplane pressure in the interarm region above
that provided by the weight of the material above.

\end{abstract}


\section{Introduction.}

Spiral structure is an astounding characteristic of disk galaxies.
Yet, the nature of the spiral structure of our home galaxy is far
from well understood.  The generally accepted model, originally
discussed by \citet{lin60}, \citet{lin61} and \citet{lin64},
involves a density wave
resulting from a global gravitational mode, triggered either by a
internal instability or some driving element (like a bar or an
interacting external galaxy).  Another proposed model describes the
spiral structure as a self-propagating wave of enhanced star formation
\citep{mue76}.  In either case, the models show the importance of
the gaseous disk in the global spiral structure phenomenon.

The role of the gas in the spiral structure has been exploited
repeatedly in order to trace the spiral arms, both in the Milky Way
and in external galaxies.  Those efforts have included \ion{H}{1}
\citep{oor58}, CO clouds \citep{dam86} and dust \citep{dri00} as
tracers.  Since the spiral arms show an enhanced star formation
rate, Pop I objects can also be used as tracers.  The model
presented by \citet{geo76}, which featured a four-arm spiral traced
by galactic \ion{H}{2} regions, is frequently cited.  Nevertheless,
it is not unusual to observe differences in the spiral structure
when different tracers are used.  One example is NGC 2997, in which
one clearly defined arm in optical light is absent in infrared light
\citep{blo94}.  In the Milky Way Galaxy, \citet{dri00} found that
the dust emission is better fit by a four-arm spiral, while a
two-arm spiral fits the stellar emission \citep[presents a
review]{val02}.

As a result of the coupling between the gaseous disk and the
interstellar magnetic field, the latter also is influenced by (and
influences) the spiral pattern. Since the magnetic fields are
``illuminated'' by the cosmic rays spinning along the field lines,
the so generated synchrotron emission is a direct probe of the
field. In external galaxies, the total (polarized + unpolarized)
synchrotron emission tends to follow the spiral pattern, although a
displacement between the positions at which the optical and
synchrotron emission peaks is not unusual.
In addition, analysis of the polarization direction of the 
synchrotron emission shows that the magnetic field is usually
aligned with the spiral arms \citep{bec96,bec02}. Then, it can be
argued that the large scale magnetic field plays a significant role
in the formation of spiral structures in disk galaxies.

\citet{mar98} explored what effect a strong galactic magnetic field
would have in the vertical structure of the spiral arms. The main
difference between their models and previous work was the inclusion
of a thicker, higher pressure ISM. The necessity of such environment
was pointed out by the fact that some components of the gaseous disk
were observed to have scale heights of the order of kpc
\citep{rey89, edg89}, much larger than those previously considered.
The support necessary for the weight of that gas is larger than the
thermal pressure observed in the midplane, leading to the conclusion
that non-thermal pressures must dominate the vertical hydrostatics of
the galactic disk, and that the pressure scale height might be much
larger that the density scale height \citep{bou90}.
When those pressures significantly
increase the effective $\gamma$ of the medium, the gas flowing into
the spiral arms shows a combination of a shock and a hydraulic jump
at the position of the spiral arms, which induced large vertical gas
motions. Such behavior appears to have been observed in NGC 5427
\citep{alf01}.

In an earlier work \citep[Paper I from here on]{gom02}, we extended
\citet{mar98} analysis to three dimensions and included a large fraction of
the galactic disk.
It is the purpose of this work to further explore those results
and aim more directly at the Milky Way structure.
The paper is organized as follows:
Section \ref{section_setup}
summarizes the results of Paper I, presents the initialization
of the simulations we present here, and describes the basic features
of the flows at an age of $800 \Myr$;
Section \ref{section_bfield} describes the magnetic field
structure, the synchrotron emission, and the field-density relationships;
Section \ref{section_hydrostatics} explores the elements
of the support of the disk and the vertical hydrostatics
near the solar circle;
Section \ref{section_periodicity} explores periodicity and
normal frequencies in the simulation;
Section \ref{section_circulation} describes the large scale
motions encountered in the simulation and evidence for circulation;
and Section \ref{section_conclusions} presents our conclusions.


\section{The simulation setup.}\label{section_setup}

Details of the setup are provided in Paper I.
Here we present only an overview, mentioning the differences
from the cases presented there.

We performed three-dimensional MHD simulations
using the code ZEUS \citep{sto92a, sto92b, sto92c} to model the
ISM response to a spiral gravitational perturbation.
The gas starts in vertical and radial
hydrostatic equilibrium, following
the circular orbits defined by the
background gravitational potential from model 2 in \citet{deh98},
the pressure gradient (thermal plus magnetic), and the
magnetic tension.
The magnetic field is azimuthal, with a strength
defined by the relation:

\begin{equation}
p_B=p_M \frac{n}{n+n_c} \dyn\cm^{-2},
\label{mag_pres_eq}
\end{equation}

\noindent
where $p_B$ is the magnetic pressure,
$p_M= 3.5 \times 10^{-13}$ and $n_c=0.04 \cm^{-3}$.
The equation of state for the initialization is isothermal
with a temperature $T = 10^4 \degK$.
As shown in Paper I, by defining the density profile in the
midplane (in our cases,
an exponential with a scale length of $4 \kpc$\footnote{
This small radial scale length for the midplane density
was used to avoid difficulties with very low densities at
small $r$ and large $z$.},
with $n = 1.11 \cm^{-3}$ at $r=8\kpc$), all the 
equilibrium variables are easily found.

Again, the hydrostatics, the magnetic field geometry, and
the density-magnetic pressure relation are only enforced in
the initialization.
All of them are altered considerably by the MHD evolution.

The simulations were carried out in a cylindrical coordinate
system.
The grid spans from $3 \kpc$ to $11 \kpc$ in the radial
direction, 0 to $1 \kpc$ in $z$ and 0 to $\pi/2$ in azimuth
for the four-arm models (0 to $\pi$ in two-arm models).
Both $r$ and $z$ boundaries are reflective, while periodic
boundaries are set in the azimuth.

\begin{figure}
\plotone{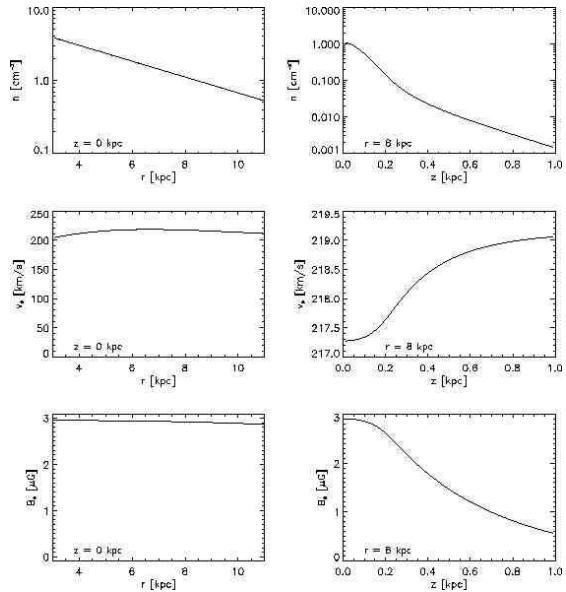}
\caption{
Density, circular velocity and magnetic field in the
initialization.
Left panels show midplane quantities versus radius.
Right curves show vertical distributions at $r=8 \kpc$.
The gas is initially in hydrostatic equilibrium, with the
midplane density
decaying radially with a scale length of $4 \kpc$.
The velocity curve is basically flat, increasing only
marginally with $z$ to accommodate the radial magnetic tension.
The magnetic field is fully azimuthal only in the initialization.
}
\label{initial}
\end{figure}

This initialization yields the distribution presented in Figure
\ref{initial}.
With slightly different parameters
($p_M=1.75 \times 10^{-12}$ and $T = 5700 \degK$),
the vertical density at the solar circle would closely resemble
the one described in \citet{bou90}.
We chose to use a lower magnetic pressure because we expected
that the accumulation of the gas in the spiral arms
would increase the magnetic field to values closer to those observed.
Also, we expected that the gas dynamics could possibly
generate a random component of the field; this did not occur.
There are several possible reasons: low spatial resolution,
numerical diffusivity, a low effective Reynolds number,
a near laminar flow, or a combination of all.
We chose to use a higher temperature to help offset the smaller
magnetic pressure (this was necessary because, without the
extra pressure, the density drops very fast with $z$, generating
numerical problems). 
It should be noted that our relatively high ``thermal'' pressure
is a rough approximation to the pressures provided in fact by
the random field, cosmic rays, subgrid turbulence, and the
comparatively weak true thermal pressure.
Since this pseudo-thermal component is isothermal (except at high
density, see discussion below), it did not add the
stiffness to the medium that the random field or other
components would have.
As a result, we likely under-represent the jump-like
character of the flow and the associated vertical and turbulent
motions.

The rotation curve versus $r$ is nearly flat,
increasing only slightly with $z$.
From Equation \ref{mag_pres_eq}, the Alfv\'en velocity
increases with distance from the midplane, to an asymptotic value
of about $30 \kms$.

\begin{figure}
\plotone{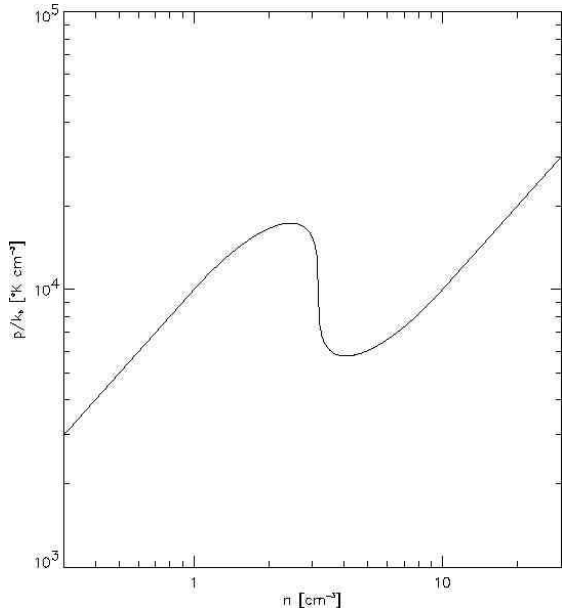}
\caption{
Equation of state used after the simulation started.
Below $n=1 \cm^{-3}$ and above $n=10 \cm^{-3}$, the gas
is isothermal with $T=10^4 \degK$ and $T=10^3 \degK$,
respectively.
In the middle, the temperature falls, mimicking abrupt cooling,
and generating a thermally unstable regime.
}
\label{two-phase}
\end{figure}

After the simulation started, we changed the equation of state
in order to facilitate the condensation of the gas into the
spiral arms.
Gas with $n < 1 \cm^{-3}$ behaves isothermally with $T=10^4 \degK$;
at higher densities, the temperature of the gas
is forced down, generating a thermally
unstable regime, up to a density of $10 \cm^{-3}$, above which
the gas behaves isothermally again, with $T = 10^3 \degK$,
effectively forming a two-phase medium (Figure \ref{two-phase}).
This change will have the biggest effect near the midplane,
where thermal pressure dominates the magnetic pressure.
The choice of $10^3 \degK$, rather than $\sim 10^2 \degK$ that one might
expect from the heating-cooling balance, was made because
this pseudo-thermal pressure represents other components that cannot
be radiated away, and because of our coarse resolution.
It is a modest attempt to sample the effects of condensation, either
due to reduced thermal pressure or self-gravity.

In contrast to Paper I,
the calculation is performed in the reference frame of the
spiral perturbation, which moves with an angular velocity
of $12 \kms \kpc^{-1}$.
The perturbation locus is a logarithmic spiral with a pitch angle of
$15 \grad$.
In radius,  the potential perturbation
varies in such a way that the implied density perturbation decays
exponentially with a scale length of $8 \kpc$.
In azimuth, the perturbation is sinusoidal,
with a peak-to-valley amplitude corresponding to
an arm/interarm mass ratio of 3.16 at $r= 8\kpc$.
In the midplane, the perturbation force is a relatively constant
fraction, $\sim 7\%$, of the average radial force.
Details of the radial and vertical shape of the perturbation
are presented in \citet{cox02}.

The perturbation is turned on linearly during the first $50 \Myr$
of the simulation.
Also, to avoid splashing against the inner radial boundary,
the perturbation forces are
not applied in the $3 \kpc < r < 4 \kpc$, and
smoothly increase to full strength in the next kpc.
Therefore, the actual useful grid extends from 5 to 11 kpc.

In summary, the differences between the models presented here
and those of Paper I are: the new equation of state, a higher
magnetic pressure in the initialization, and the shift of the
calculation to the pattern reference frame.
Also, in order to avoid some of the numerical problems
described in Paper I, we set a density floor of
$n=4.75 \times 10^{-9} \cm^{-3}$, equal to the minimum
density in the initialization.
We were consequently able to run the simulation to much greater ages.

The setup described should be susceptible to the Parker instability
\citep{par66}.
Nevertheless, we do not expect to see the interchange mode of the instability
due to our low resolution, while the undular mode is expected to be
quenched by both our lack of resolution and the flow we are
modeling.
In order to test this expectation, we restricted the computational
domain to a two-dimensional grid along the solar circle, and set up
an atmosphere like the one described above.
When we turned the arm perturbation off, we observed the growth of
the undular mode of the Parker instability.
But, when the perturbation was added, the magnetic field lines bent
due to the vertical motion of the gas generated by the hydraulic jump,
and not so much by the weight of the gas.
So, any vertical perturbation in the magnetic field will be carried
along with the orbiting gas, and will be damped out when it
encounters the hydraulic jump.
Therefore, as long as the arm-encounter time is shorter than the
instability growth time scale, the undular mode of the instability
should not be present.


\begin{figure}
\plotone{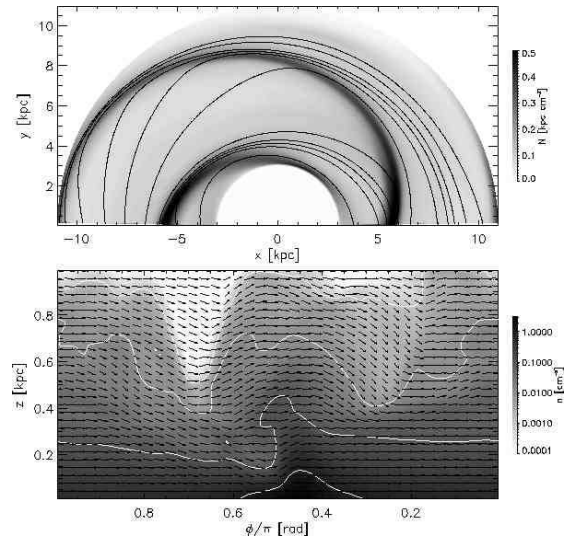}
\caption{
Structure of our two-arm case after $800 \Myr$.
In the upper panel, the grayscale shows the half-disk column
density, and the lines show the integrated pattern frame
velocity field in the midplane.
In the lower panel, the grayscale show a density cut along
a cylindrical surface at $r=8 \kpc$ with contours at each
decade, down from $n=1 \cm^{-3}$.
The arrows show the velocity field component parallel
to that surface, in the pattern reference frame.
Since ours is a trailing spiral, the gas flows clockwise
in the upper panel, and from left to right in the lower one.
In the bottom panel, the gravitational perturbation potential
minimum occur at $\phi/\pi = 0.5$.
}
\label{two_arm}
\end{figure}

\begin{figure}[b]
\plotone{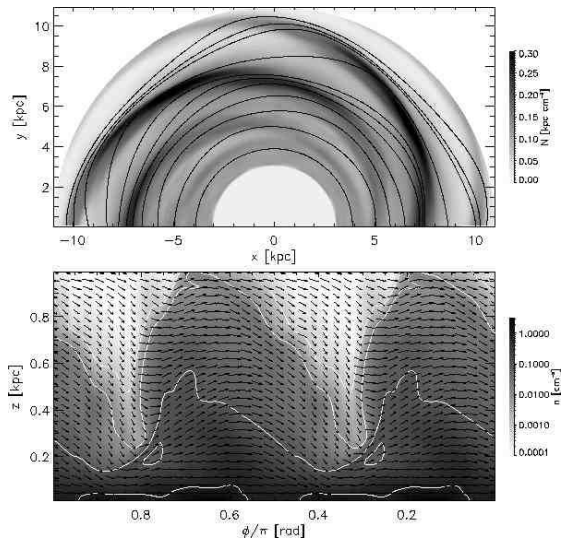}
\caption{
Same as Figure \ref{two_arm}, for our four-arm model.
In the bottom panel, the gravitational perturbation potential
minima occur at $\phi/\pi = 0.35$ and $0.85$.
}
\label{four_arm}
\end{figure}

\subsection{General behavior of the simulations.}
\label{behavior_sec}

Figures \ref{two_arm} and \ref{four_arm} show the structures
of our two standard models (two and four arms) after 800 Myr.
(Unless otherwise noted, we report the state of the
simulations at this time.)
In both Figures, the grayscale in the upper panel
shows half-disk column density;
the solid lines show the integrated pattern frame velocity field in the
midplane.
The grayscale in the lower panels shows a density cut
at a cylindrical surface with $r=8 \kpc$.
The arrows represent the velocity field in the pattern
reference frame.
The length of the
arrows is proportional to the total velocity in the $r=8 \kpc$
surface; but, since the vertical axis is greatly stretched
(in reality, the aspect ratio of the grid is
$r \Delta \phi:\Delta z \approx 24:1$)
the individual components of the velocity are also stretched so that
the arrows point in the corresponding direction with respect
to the density structures.
As these are models of trailing spirals, the gas rotates
clockwise in the upper panels, and from left to right in the lower
ones.

In Paper I, we showed that, as the gas enters the arm,
a combination of a shock and a hydraulic jump is formed.
The extra stiffness the magnetic field adds to the ISM
makes the gas jump over the obstacle the gaseous spiral arm represents.
The gas shoots up before the arm, forming an forward
leaning shock,
and falls back down after the arm, generating a secondary
shock.
In isothermal cases with no magnetic field (Paper I),
there is much less vertical motion,
the forward shock is nearly vertical,
and there is no secondary shock.


\begin{figure*}
\plottwo{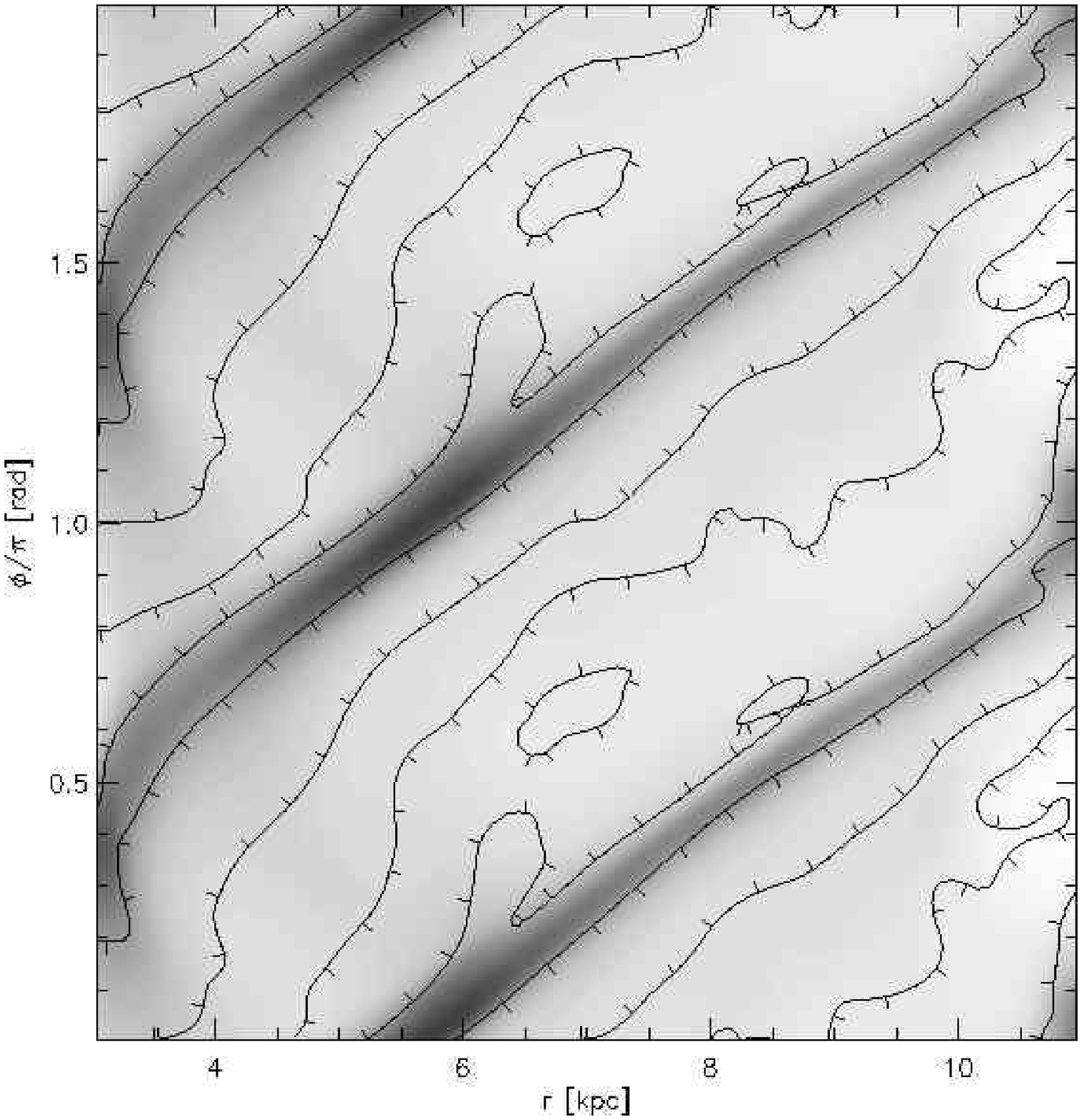}
        {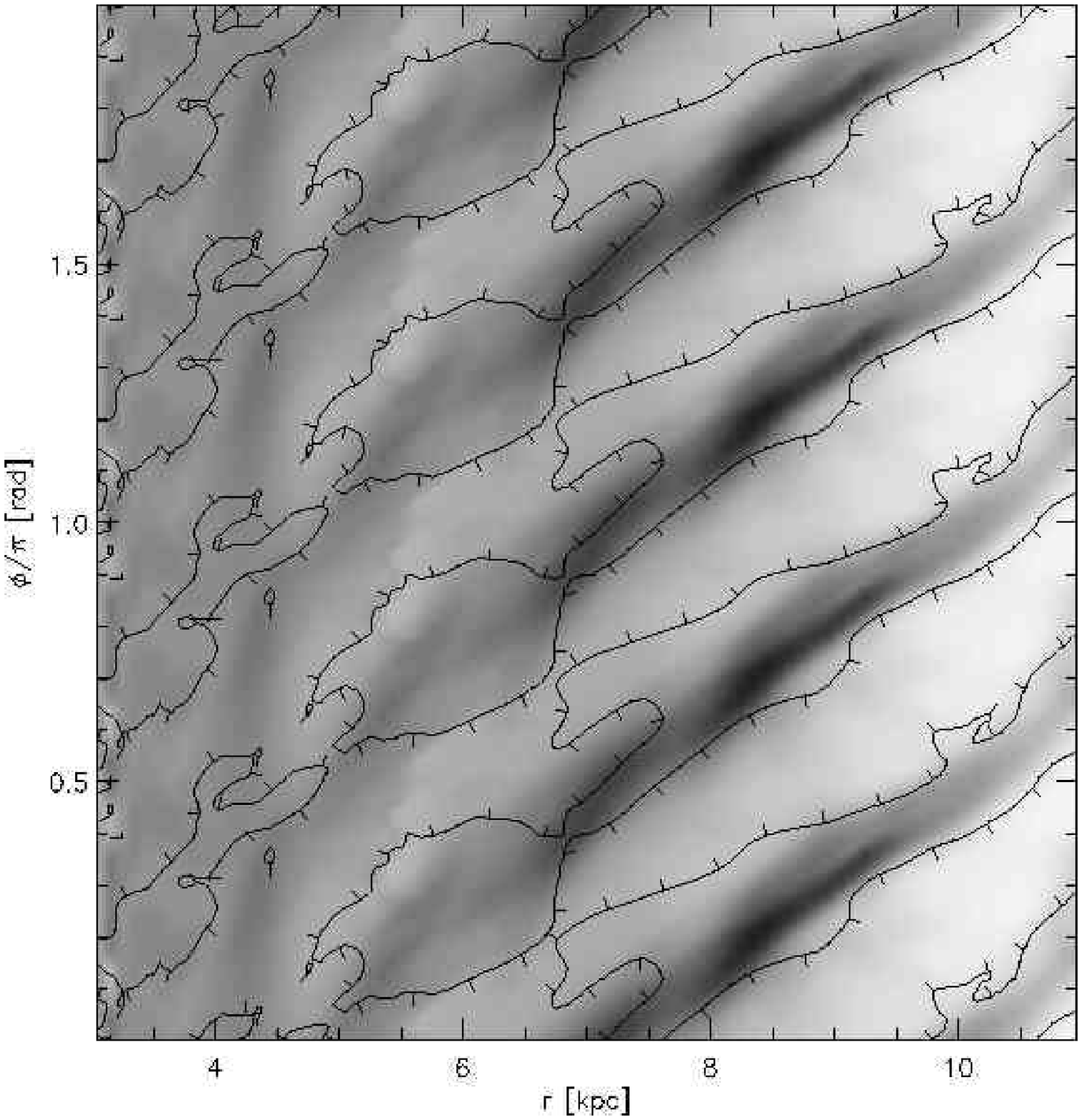}
\caption{
Integrated vertical mass flux for
the two-arm case (left panel) and the four-arm case (right panel).
In both panels, the grayscale shows the column density of
the simulation.
The contour shows the zero-level of the vertical flux, with the
dashes pointing in the downflow direction.
The gas flows up at the arms and down in the interarms.
The two-arm case has a second up-down structure in the
interarm region.
In both cases, there is a tendency for frequency doubling
around $r=7\kpc$.
Gas flows down from the top.
}
\label{total_flux}
\end{figure*}

\subsubsection{Two arm model.}

In the two-arm case, when the gas falls behind the arm,
it bounces back up generating a high $z$ interarm structure that
mimics another arm.
The high $z$ material then falls again before reaching
the next arm.
This structure does not show up in the column
density, but it is clear in both velocity and density in
the lower panel of Figure \ref{two_arm}.
It is explored in the left panel of Figure \ref{total_flux},
where the contours show the zero-level of the
integrated vertical mass flux ($\int \rho v_z ~dz$), with the
dashes pointing in the downflow direction.
The column density is shown in grayscale.
The downflow region is broader and happens at lower
densities than the upflow.
Also, there is a second upflow region downstream of the arm,
that mimics the flow structure around the arm.
A powerful bounce therefore appears to be a mechanism that
might double the number of gaseous arms relative to the stellar
arms, as is apparently observed in some galaxies.

Generally speaking, in contrast with the model of \citet{rob69},
the gaseous arms sit downstream from the
perturbation minimum, but follow a tighter spiral (with a
pitch angle of about $13 \grad$).
We have not yet discovered the reasons behind either the pitch
angle difference (which appears in both 2D and 3D models) or
the phase difference relative to \citet{rob69} result.


\subsubsection{Four arm model.}\label{four_arm_subsection}

The four-arm case (Figure \ref{four_arm})
has a similar structure to the two-arm
one, but presents a big difference: when the gas falls
behind the arms, it does not have enough space to complete
a bounce and fall again
before it encounters the next arm.
The downflow does lead to vertical compression ahead of the
next arm, but not to a doubling of the vertical velocity
pattern.
The leading shock is also more vertical, especially at
higher $z$.
The right panel in
Figure \ref{total_flux} shows the vertical mass flux for
the four-arm case.
Here, the upflow region is broader than in the two-arm
case and, as noted above, there is no
obvious interarm bounce structure.
(Both two and four-arm models show higher numbers of features
in narrow radial ranges, notably around $r=7 \kpc$.
This may be due to a resonance between vertical oscillations
and arm passage time, and is associated with broadening or
even doubling of the arm structure.)

\begin{figure}[b]
\plotone{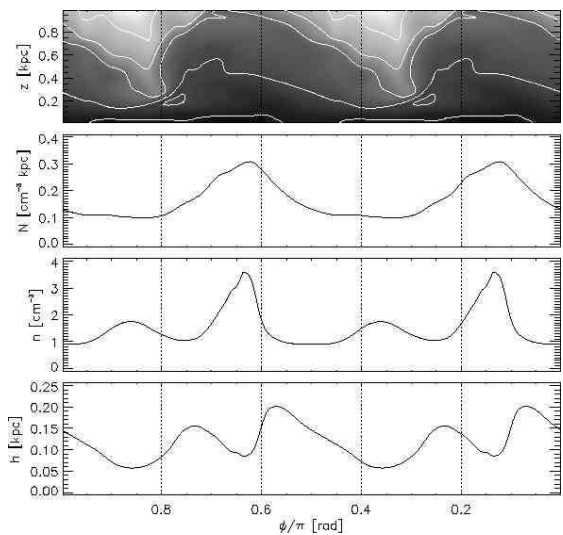}
\caption{
Comparison of the azimuthal extension of the arms for
$r=8\kpc$ in the four-arm model.
Starting from the top,
the first panel shows the density on a surface at constant
radius, with contours at every decade, starting at $1 \cm^{-3}$.
The second shows the half-disk column density,
the third shows the midplane density, and the last panel
shows the disk scale height, defined as the ratio
of the column to midplane densities.
}
\label{swollen}
\end{figure}

It is not clear how to separate the arm and the interarm
regions.
We illustrate this in Figure \ref{swollen}.
As shown in the top panel (a reproduction of
the lower panel of Figure \ref{four_arm}),
the high $z$ structure associated with a
spiral arm is very wide, as traced by the density contours.
In fact, the region over which the disk thins
(the contours reach low $z$) is very narrow.
On the other hand, the column density (second panel from the
top) also has clearly defined arms, but they cover a smaller
region in azimuth than the high $z$ density structure.
The arms are even narrower when we look at the midplane
density (third panel from the top).
Notice that,
even if the gas flow does not bounce in the interarm,
the downflow compresses the disk and
generates an overdensity near the midplane
before it encounters the next arm.

A frequently used measure for the thickness of the disk is the
ratio of the column density to the midplane density, which
is an estimate of the scale height.
Examination of the top panel of Figure \ref{swollen}
leads to the idea that the disk swells at the arms.
But, since the arm  characteristics have
different extents along azimuth,
the estimated scale height (bottom panel) has a local minimum
in the middle of the arm.
In the leading and trailing end of the arm, the scale
height increases, as expected.
This difference in the behavior of the column and midplane densities
causes the scale height defined in this way
to behave counter-intuitively.

\begin{figure}[b]
\plotone{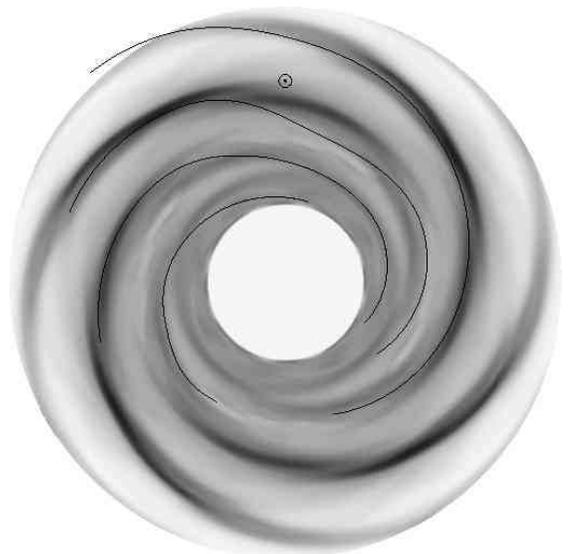}
\caption{
Column density of our four-arm model, compared
with the positions of the Milky Way's arms, as traced
by \citet{tay93}. The position of the Sun with respect to the
modeled arms is marked.
}
\label{match_arms}
\end{figure}

In Figure \ref{four_arm}
the arms seem to be discontinuous, with a break
at about $r=7.5 \kpc$.
Each of the sections follows an even tighter spiral
(with $11 \grad$ pitch angle) than the perturbation ($15\grad$)
and the overall locus of the arm ($13 \grad$).
The locus pitch angle
has the advantage that the resulting arms are a good
approximation to the Milky Way's spiral arms as
traced by \citet{tay93}.
Figure \ref{match_arms} shows the column density
of our four-arm model and the locus of the arms in the
aforementioned work, scaled so that the distance from
the Sun to the Galactic center is $8 \kpc$.
This agreement is extensively used in \citet{gom04},
where we generate synthetic observations using this model,
for comparison with real Milky Way data.


\section{Magnetic field.}\label{section_bfield}

Figures \ref{two_arm} and \ref{four_arm} show that,
as material in the midlpane
approaches the arms, it moves radially outward,
then shocks and follows a path nearly parallel to but soon outside
of the arm, moving radially inward in the process.
As we discuss below, the post shock inward flow at
the midplane is smaller than the outward preshock excursion,
and there is a slight migration outward balanced by inflow
at higher $z$.

\begin{figure}
\plotone{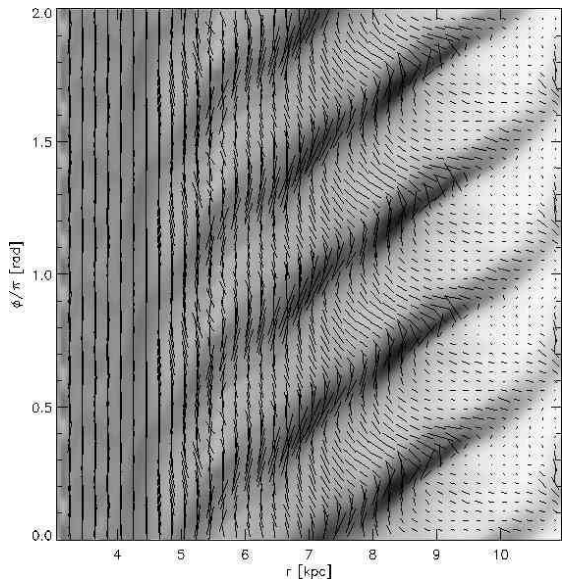}
\caption{
Pitch angle of the magnetic field in the midplane
for the four-arm case and its relation with column density.
The grayscale shows the column density of the gas, while
the lines show the direction of the magnetic field,
with length proportional to the intensity in that plane.
As with Figure \ref{total_flux},
the gas flows down from the top.
}
\label{v.b1}
\end{figure}

\begin{figure}
\plotone{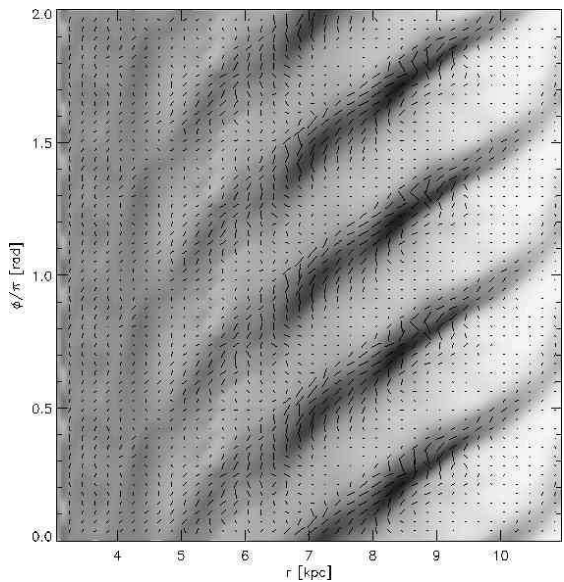}
\caption{
Same as Figure \ref{v.b1}, for $z=0.5 \kpc$.
The length of the lines (proportional to the field strength)
have the same normalization as in Figure \ref{v.b1}.
Patches where the field strength is very small must lie in regions
where the higher $z$-lower field material has moved downward
to this level.
Conversely, strong fields are present where the stronger field
from below has been advected upward.
A sudden jump from one to the other occurs on the leading edges
of the arms for $r>7\kpc$.
}
\label{v.b2}
\end{figure}

The magnetic field is carried along with this flow, and
therefore has a radially outward component ahead of the arms
and a radially inward component just outside of them.
On Figure \ref{v.b1}, these radial
excursions of the field lines are plainly seen.
At the locations of the arms, the field is roughly parallel
to the arm locus.
Between the arms, the field pitch changes from somewhat less
positive than the arm pitch, to circular, to negative
pitch as it approaches the next arm.

At higher $z$ (Figure \ref{v.b2}), both the velocity
pattern and field structure are somewhat less regular.
As expected from the gas flow,
the largest vertical component of the magnetic
field is found just before and after
the arms.
The vertical field can be as large as $0.4 \microG$,
but $0.2 \microG$ is a more typical value.

The two-arm case (not shown) has a much smoother magnetic field.
It can also show negative pitch in the interarm region,
but it does not change as abruptly as in the four-arm case.
Also, the vertical field is weaker, being prominent only on
the leading edge of the gaseous arms.


\begin{figure}
\plotone{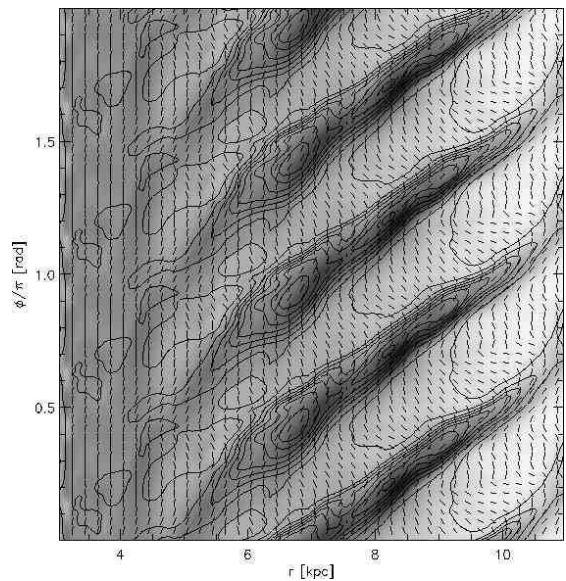}
\caption{
Synchrotron emission of the four-arm model.
Grayscale shows the gas column density, while the contours
show the intensity of the emission.
Dashes show the direction of the magnetic field as inferred
from the polarization.
}
\label{synch}
\end{figure}

\subsection{Synchrotron emission.}

Owing to our missing the random field component, a synthetic
synchrotron map cannot represent the emission actually expected
from a galaxy, but it could provide a reasonable view
of the polarized fraction.
Figure \ref{synch} shows our calculated total emission,
for our four-arm case,
as viewed from outside the galaxy, with lines showing the
directions of the apparent field (perpendicular to the
$E$ field polarization of the integrated emission),
assuming no Faraday rotation.

At any given distance above the plane, the emissivity is given by:

\begin{equation}
\varepsilon_{tot}(r,\phi) \propto n_e(r,z) B_\perp^{(p+1)/2},
\end{equation}

\noindent
where
$B_\perp$ is the component of the field
parallel to the plane of the galaxy,
$p=2.5$ is the spectral index
of the electron cosmic ray distribution,
$n_e(r,z)=\exp(-r/r_{cr}-z/z_{cr})$ is the spatial distribution
of the electron cosmic rays,
$r_{cr}=13 \kpc$, and $z_{cr}=2.5 \kpc$
\citep{fer98}.
This emissivity is integrated along a line of sight perpendicular
to the galactic plane.
The polarization is calculated by estimating the polarized emission
parallel and perpendicular to $B_\perp$,

\begin{eqnarray}
\varepsilon_\parallel &=& \frac{1+\Pi}{2} \varepsilon_{tot} \cr
\varepsilon_\perp     &=& \frac{1-\Pi}{2} \varepsilon_{tot},
\end{eqnarray}

\noindent
where $\Pi=(p+1)/(p+7/3)$ is the degree of polarization
of the emissivity.
Given the angle $\alpha$ and the intensities
$I_{tot}$, $I_\parallel$ and $I_\perp$ at a certain $z$,
the intensities and parallel direction
at $z+dz$ are given by:

\begin{eqnarray}
\alpha' &=& \alpha \nonumber \\
 & & + \frac{1}{2} \atan\left[
  \frac{(\varepsilon_\perp - \varepsilon_\parallel) dz
        \sin[2(\beta-\alpha)]}
       {(I_\perp-I_\parallel)+
        (\varepsilon_\perp - \varepsilon_\parallel) dz
        \cos[2(\beta-\alpha)]}\right] \nonumber \\
 & & \\
I'_{tot} &=& I_{tot} + \varepsilon_{tot} dz \\
I'_\parallel &=& I_\perp+\varepsilon_\perp dz
              -(I_\perp-I_\parallel)\cos^2(\alpha'-\alpha)
                                                \nonumber  \\
          & & -(\varepsilon_\perp-\varepsilon_\parallel) dz
               \cos^2(\alpha'-\beta) \\
I'_\perp &=& I'_{tot}-I'_\parallel,
\end{eqnarray}

\noindent
where $\beta$ is the direction parallel to the
magnetic field at $z+dz$.
Notice that $\alpha'$ is not the direction of polarization
nor the direction of the magnetic field, but the direction
of the inferred B-field, as traced by the synchrotron emission.

In Figure \ref{synch} we can see that the interarm turning
of the
B-field happens around the area of minimum emission, which is
consistent with the picture of being a consequence of stretching the
field.
The higher intensity regions tend to be slightly upstream from
the arm.
The degree of polarization is quite large (about 70\%).
This is expected, since our resolution does not allow us to model
the random component of the field, that might depolarize
the emission.
Only a shift in the direction of the B-field with $z$ reduces
the polarization from the $72.4\%$ of the intrinsic emissivity.

For our two-arm case (not shown),
the polarization changes are not as abrupt, the arms are more
continuous (no blobs with associated  peaks of synchrotron
emission), and the difference in the positions of the peaking of
emission and column density is smaller.

Figure \ref{synch} is to be compared with the maps from
\citet{bec02}.
In that work, an atlas of synchrotron emission in external
galaxies is presented.
They compared the polarized and total emissions, as a way of
comparing the total (random + regular) and regular B-fields
(again, we do not have enough resolution to say much about
the random field).
They consistently found islands of synchrotron emission
and displacements between the synchrotron
and gaseous arms not so different from the ones found here.
We did not find magnetic arms without a density counterpart,
which have been occasionally found in polarized emission
observations and some simulations \citep[for example]{els00}.

Galaxies should have a lot of random field in the spiral arms,
generated by both the turbulence of the jump and stellar
feedback, but the post arm flow should stretch the field to
greater regularity in the interarm region.  
To the extent that this is true, we would expect the mean convected
field to follow the pattern just described.  
Some observations \citep{han99} suggest that in some galaxies
there are magnetic arms,
regions of enhanced polarized synchrotron emission not
associated with the gaseous arms.  
Our finding
that in two-arm spirals, with their longer interarm transit,
the flow tends to bounce and generate an arm-like structure in
the interarm region could also be related to this phenomenon.  
A detailed exploration of these points is outside the scope of
the present work.


\subsection{Magnetic field-density relation.}

The relationship between magnetic field strength and density
in the interstellar medium is an interesting diagnostic of the
behavior of the flow.
To the extent that the various
components of the ISM are all just the current states of
material that samples all states, so that there is no
intrinsic relationship between density and flux, the
relationship should distinguish the various dynamical
possibilities.
If the field were small enough to be
dynamically insignificant, compressions on average should be
three dimensional, which with flux freezing would yield
$B \propto n^{2/3}$.
If the field were so strong that it dominates the
pressure, compressions would be possible only along the field,
and there would be no dependence of field strength on density,
except in dynamically active environments.
Measurements of the field strength in the midplane, at average
and lower densities seem to find no dependence of $B$ with $n$
\citep{hei01}.  
At densities high enough that local dynamics or self gravity
is important, higher field strengths (and higher total
pressures) are found \citep{cru99}.
At the lower densities, some
measurements provide only upper limits to local region fields,
but the strong fluctuations in pulsar dispersion measures and
the overall synchrotron emissivity seem to imply that the rms
field is several microgauss.
It would appear that most of the
volume, even at low density, must have fields of this order
(see discussion in Vall\'ee, 1997).

Our simulation can be explored for its $B(n)$ relationship,
though some care is required in interpretation. 
Two factors must be kept in mind.
First, we initialize our run with a very
strong dependence of $B$ on $n$, as given by Equation
\ref{mag_pres_eq}.
It is
almost certain that far from the plane of the Galaxy, not only
is the density low, but so is the magnetic field.
Our initialization is a specific rule for this relationship, and
determines the static vertical structure.
We observe, however,
that even in our relatively laminar flow,
material makes large vertical excursions in its response to
the arm disturbance (see Section \ref{section_circulation}).
Thus, the material found at any given height will have
a wide spread in initial heights,
and therefore, a wide spread in initial $B(n)$.
This is more true further from the
plane, but should not be discounted close to the plane as well.

The second point is that because we do not have a significant
random field, it may be easier for material to move along the
field than in the actual Galaxy.
This is not necessarily true
if, for example, most of that motion occurs in interarm
regions where the field is less random (as discussed above).  
It is also possible that in the actual Galaxy, a considerable
amount of motion along fields is driven at small scales and
high densities by stellar feedback in the arms, and then
frozen in as the material and field are stretched out between
the arms, the field returning to its pressure defined
equilibrium, with whatever distribution of density is left
from the small scale dynamics.

In short, the results we are about to present are suggestive,
but cannot be assumed to represent reality very precisely.

\begin{figure}
\plotone{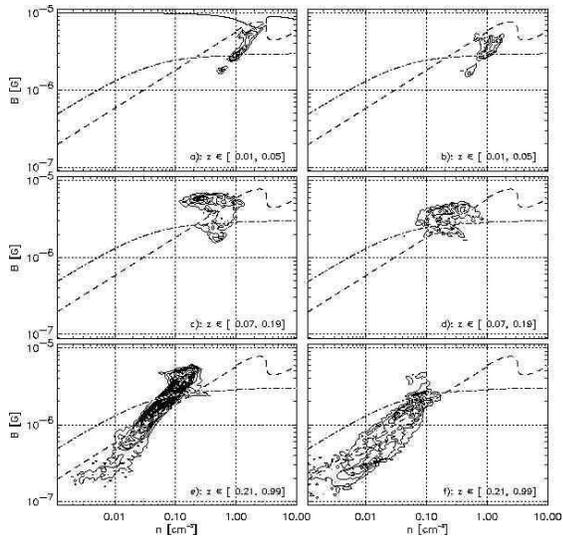}
\caption{
Distribution of density and magnetic field strength between
$r=7$ and $9\kpc$ for the four-arm model.
The left column corresponds to gas in and near the spiral arms,
while the right column corresponds to interarm gas.
The solid line in panel $a$ traces the amount of magnetic field
needed
for a total pressure of $4 \times 10^{-12} \dyn \cm^{-2}$.
The dashed line in all panels shows the locus of equal magnetic
and thermal pressures.
The dot-dash line follows the magnetic pressure-density
relation used in the initialization.
}
\label{mag_dens_arms}
\end{figure}

Owing to our strong $B(n)$ in the initialization, we restrict
our presentation to material found between 7 and $9 \kpc$, and
present three ranges in $z$, close to the plane, intermediate,
and high, the boundaries between them being at 50 and $200 \pc$.  
We also separate the material into two equal halves (in
space), one containing the arms, the other not.
The results
are shown in the six panels of Figure \ref{mag_dens_arms}.
The contours are
of volume containing $B(n)$.

There are also several lines on these diagrams, one showing
the $B(n)$ of the initialization, one showing the locus of equal
magnetic and thermal pressures, and one (shown only on the
midplane arm plot) showing a line of constant total pressure
of $4 \times 10^{-12} \dyn \cm^{-2}$.

There are several things to notice.
First, at no height does
the initial trend of $B(n)$ represent the trend of the data,
though in every diagram, significant amounts of material are
found along parts of the initialization.
Next, there are
three characteristic features found in the plots, sections
where the distribution has a major horizontal feature,
sections where there is a major feature or slope somewhat
steeper than the initialization, but lower than the slope
(one) expected from purely transverse compression, and in the
midplane, a feature at the highest densities which has a
concave upward boundary.

The horizontal segments, most noticeable at the intermediate
heights, are exclusively found in regions in which the
magnetic pressure is higher to significantly higher than the
thermal pressure.
Our flow thus seems to reproduce the
expectation that in such regions the dominance of magnetic
pressure is accompanied by flows parallel to it.
The vertical
width of the horizontal segment should then be associated with
the spread in pressures found between 50 and $200 \pc$ in the
central pair of panels.
The width is roughly a factor of
1.4 in $B$ and therefore 2 in pressure, more or less consistent
with the pressure variation with height presented in Section
\ref{section_hydrostatics}.
The fact that this behavior is not well represented in the
midplane is likely due to our initialization with dominant
thermal pressure at the midplane densities.
Notice that in
our simulation, the densities within $50 \pc$ of the midplane are
never as low as those expected for intercloud regions in the
Milky Way.

The inclined segments have $B \propto n^{0.7}$ or so, roughly as
expected from isotropic compression, a little steeper than the
initialized relationship.
Instead of showing isotropic
compression, it could be a mixture of the initialization and
the $B \propto n$ expected from purely transverse compression.
We do not therefore give definitive reasons for this behavior,
but note that it is always found in material whose magnetic
pressure is less than or comparable to its thermal pressure.
It is
in the regime for which isotropic compression is plausible.

The third feature, the concave upward behavior at the highest
densities close to the plane, appears to be a consequence of
our thermal equation of state.
This behavior occurs at those
densities for which the temperature is rapidly decreasing with
increasing $n$,
requiring a higher value of $B$ to have the same total pressure.
Notice the similarity of the locus to the constant pressure
line superimposed.


\section{Vertical hydrostatics.}\label{section_hydrostatics}

Various authors \citep[and references therein]{bou90}
have attempted to reconcile the apparent midplane pressure
of the Galaxy with the weight of the material above it.
In this section, we test two aspects of this procedure,
whether such equality is expected locally (for example,
in the immediate neighborhood of the Sun,
there appears to be a scarcity of material relative to the average
and one might wonder whether this should be reflected in
a lower than average midplane pressure),
and the degree to which dynamics contributes to the overall balance.

\begin{figure}[b]
\plotone{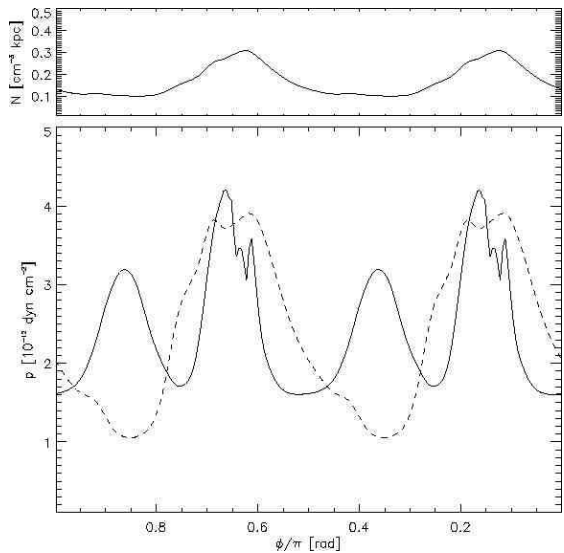}
\caption{
Comparison of the midplane pressure and gas weight for $r=8\kpc$
for the four-arm model.
In the lower panel, the solid line shows the midplane pressure,
while the dashed line shows the integrated weight.
In the places where the weight is higher than the pressure,
the downward force is decelerating the upflow
(before the arm) or generating the downflow (after the arm).
The pressure is higher in the interarm because of the kinematic
pressure due to the downflow behind the arm and the subsequent bounce.
The half-disk column density is provided in the upper panel
for reference.
}
\label{pressure}
\end{figure}

In Figure \ref{pressure}
we compare the midplane thermal plus magnetic
pressure with the integral of the overlying weight per unit
area versus phase along a line of constant galactocentric
radius of 8 kpc.
The half-disk column density is also provided for comparison.
The mean value of the midplane pressure ($2.45\times 10^{-12}
\dyn \cm^{-2}$) is in good agreement with the mean weight
($2.41\times 10^{-12} \dyn \cm^{-2}$).
But the behavior with phase shows much more interesting
structure.
The midplane pressure does have significant
variations, as does the weight, and they are correlated to
some degree.
A perfect match is not expected because material does not
evolve on constant radius, because these results are for one
specific time, during which things might be slightly out of
equilibrium, and because the pressure calculated does not
include dynamical contributions.
Still, it is possible to get a sense of the dynamical behavior
from this plot.
For $\phi/\pi = 0.8$ to 1, the column density is nearly constant, so the
weight is a measure of the vertical extent.
From left to right, the weight falls, as does the material,
raising the midplane density and pressure.
The trend reverses about $\phi/\pi = 0.85$ to upflow, increasing
weight and decreasing pressure.
This is a bounce.
Then, there is a sudden increase in pressure and weight, a shock,
that raises the pressure above the weight, though only slightly.
This is the hydraulic jump.
It sustains the upflow, roughly stabilizing the increased height.
Thereafter, the flow accelerates parallel to the plane, decreasing the
pressure, weight, and column density.
Because the effective $\gamma$ is grater than 1, the pressure drops faster
than the weight, leading to downflow, a downstream pressure increase
and weight minimum, ready to bounce again.

We have explored the vertical hydrostatics more precisely,
averaging over phase to see how the contributions to pressure
versus height are arranged.
We cannot explore the importance of magnetic tension
as \citet{bou90} did, since our model has
restrictions that limit the development of vertical magnetic
field, namely, lack of spatial resolution, no cosmic rays pushing
the magnetic lines, nor an effective dynamo.
Nevertheless, we found that the large scale motions of the gas
provide a significant fraction of the support.

Consider the vertical component of the
momentum conservation equation:

\begin{equation}
\rho \left[\frac{\partial v_z}{\partial t}
  + (\mbox{\boldmath $ v \cdot \nabla $}) v_z \right] =
  -\rho g_z - \frac{\partial}{\partial z}\left(p_{th}+p_m\right),
\end{equation}

\noindent
where $v_z$ is the component of the velocity in the vertical
direction, $g_z$ is the vertical gravitational acceleration,
$\rho$ is the gas density, $p_{th}$ is the thermal pressure,
and $p_m$ is the magnetic pressure.
Here, we used the approximation proposed in \citet{bou90}, in
which the magnetic pressure is diminished by the magnetic
tension: $p_m=(B^2-2B_z^2)/8\pi = (B_r^2+B_\phi^2-B_z^2)/8\pi$.
In our model, the vertical field is not very large, and
the magnetic tension is only a small fraction of the vertical
support.
In the case of $\partial/\partial t=0$,

\begin{equation} \label{hydro_2}
p(z) = p(z_0)   +\int^{z_0}_z \rho \left(g_z +
\frac{v_\phi}{r} \frac{\partial v_z}{\partial \phi} +
v_r \frac{\partial v_z}{\partial r} + 
v_z \frac{\partial v_z}{\partial z}\right) dz.
\end{equation}

\begin{figure}
\plotone{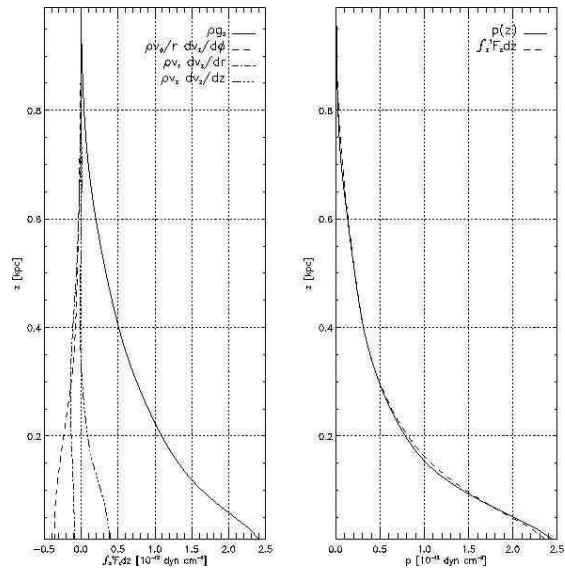}
\caption{
The left panel shows the contributions of the different force
terms to the vertical support of the disk.
Notice that the vertical kinematic pressure works in the same
direction as the gas weight.
The right panel compares the pressure [thermal + magnetic, in the
\citet{bou90} tension approximation] and the vertical forces.
When the convective terms are considered, the disk
is very close to vertical equilibrium at this epoch.
}
\label{pres_terms}
\end{figure}

\noindent
In Figure \ref{pres_terms}, the left panel shows the
four different integrated terms in the right side of Equation
\ref{hydro_2}, and the right panel
compares the left and right hand sides of the equation.
In that Figure,
$p(z_0)$ is the value of the thermal plus effective magnetic
pressure at the upper boundary at $z_0=1\kpc$.
All the plotted quantities are
actually the mean values along the solar circle, so that
the effect of the spiral arms is averaged out to give
a global picture.
In order to calculate the weight of the gas, the total
gravitational potential (background + perturbation) was
used, but it makes little difference, since the actual value
of the perturbation is small compared with the background
potential.

Figure \ref{pres_terms} shows that the spatially averaged
disk atmosphere is actually very close to hydrostatic
at the $t=800 \Myr$ time shown.
Although most of the pressure is balanced by the weight of the
gas, the convective terms have a significant influence,
specially at intermediate $z$.
As expected, the first two convective terms work
into supporting the weight of the gas, generating
kinematic pressure.
But the fourth term, the vertical kinematic pressure
$\int \rho v_z (\partial v_z/\partial z) dz$,
has the same sign as the gas weight.
This is because
its average is biased toward the highest vertical velocity regions.
Such regions are the downflow behind the arms,
where $v_z$ is negative and becoming more negative with
increasing $z$.
As hinted by Figure \ref{pressure},
the midplane disk pressure has to decelerate
the downflow in addition to supporting the weight of the gas.

At different times, the lines showing the weight of the gas
(dashed line in the right panel of Figure \ref{pres_terms})
and pressure (solid line) oscillate only slightly around each
other, suggesting that the vertical hydrostatics described
above is not atypical.


\begin{figure}
\plotone{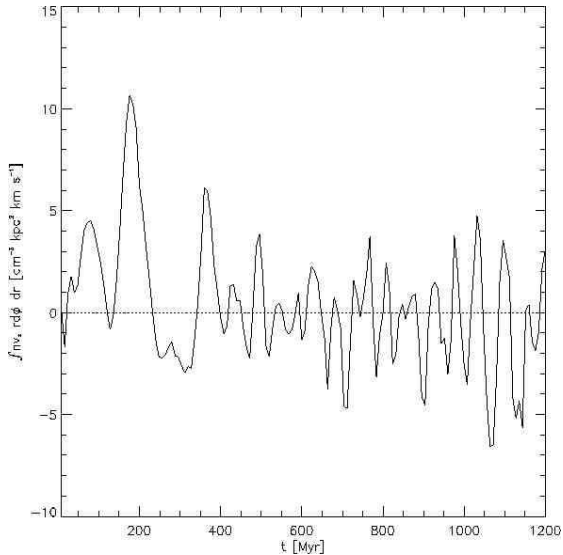}
\caption{
Vertical mass flux integrated over the plane,
at $z=210 \pc$, as a function of time.
}
\label{mass_flux}
\end{figure}

\begin{figure*}
\plottwo{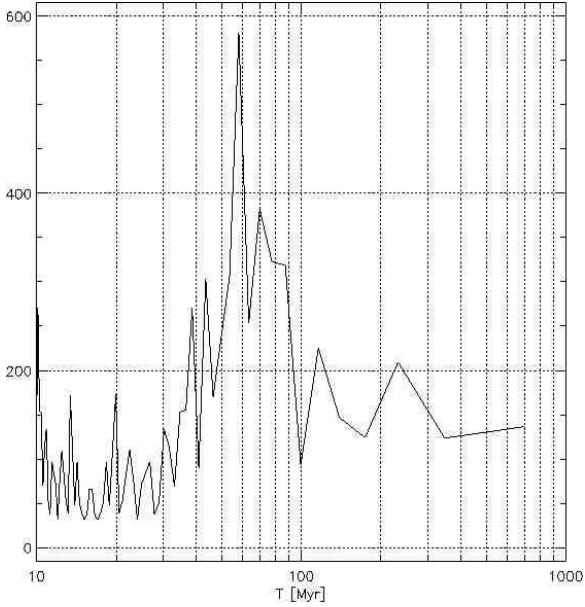}
        {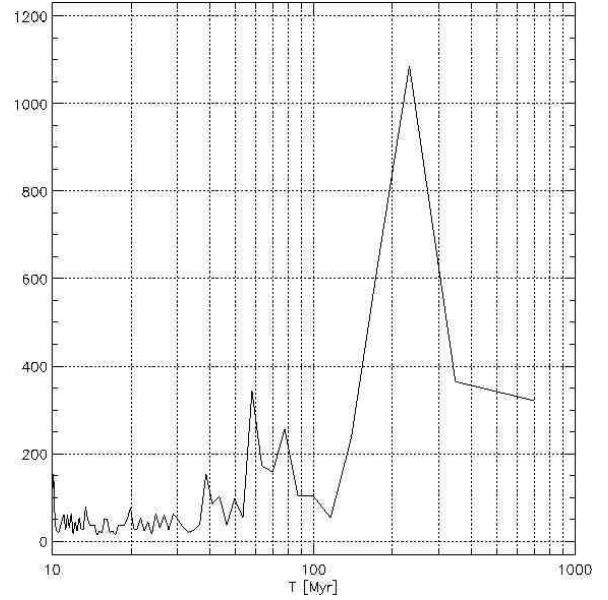}
\caption{
Fourier analysis for the vertical
mass flux at $z=210 \pc$ for the four-arm case.
When we integrate over the plane, a strong peak at
$T \sim 60 \Myr$ appears (left panel).
When we restrict the integration to $1 \kpc$
around $r= 8\kpc$, that peak diminishes, and another
one at $T \sim 230 \Myr$ appears (right panel).
}
\label{mass_flux_fourier}
\end{figure*}

\section{Periodicity.}\label{section_periodicity}

The simulations do not seem to be in a steady state, but rather
in a steady cyclic behavior,
in which the ``wave-like'' structure that is formed above the
gaseous arms breaks, only to be regenerated at the back of the arm
and rise again.
The periodicity of the simulation can be assessed by plotting
the vertical mass flow, at a given height, integrated
over the $r-\phi$ plane, as a function of time.
Such a plot for the four-arm case is presented in
Figure \ref{mass_flux}.
The chosen height is $z=210 \pc$, which is right across the
``breaking wave'' structure (see Figure \ref{four_arm}).
The left panel of
Figure \ref{mass_flux_fourier} shows the Fourier transform
of this vertical flux for $t>500 \Myr$.
It shows a strong peak at about $T \sim 58 \Myr$.
That peak grows smaller if we restrict the integration to only
a range of $1 \kpc$ around $r=8 \kpc$
($1 \kpc$ is the typical width of the gas' radial motions,
see Section \ref{section_circulation}).
The right panel of
Figure \ref{mass_flux_fourier} shows that case.
Notice that a second strong peak at $T \sim 230 \Myr$ appears.
Both peaks are present when we move the integration range
to $r=6 \kpc$ and $10 \kpc$, although they both
move to slightly smaller periods and
the large period peak becomes weaker than the
short period one as we go to smaller radii.

We performed a linear perturbation analysis of the vertical
motions of the disk in order to understand the origin
of these periods.
The procedure is very similar to the one followed in the
appendix of \citet{wal01}.
Consider the equation of motion in the vertical direction
for an plane-parallel isothermal atmosphere:

\begin{equation} \label{eq_motion}
\frac{\partial v}{\partial t} = 
  - \frac{1}{\rho} \frac{\partial p}{\partial z} - g,
\end{equation}

\noindent
where $v$ is the velocity, $\rho$ is the mass density,
$p$ is the total pressure (thermal plus magnetic),
and $g = \partial \Phi/\partial z$ is the gravitational
acceleration.
Consider perturbations to the (hydrostatic) equilibrium
distribution,

\begin{eqnarray}
\rho \longrightarrow & \rho_0 + \delta(z) \cr
p    \longrightarrow & p_0 + \epsilon(z) \cr
v    \longrightarrow & v(z).
\end{eqnarray}

\noindent
Substituting these into equation \ref{eq_motion}, and eliminating
the hydrostatic equilibrium condition, we get:

\begin{equation}
\frac{\partial v}{\partial t} = -\frac{1}{\rho_0}
  \left( \delta g + c^2 \frac{\partial \delta}{\partial z}
    + \delta \frac{\partial c^2}{\partial z}\right),
\end{equation}

\noindent
where $c^2=\partial p/\partial \rho$.
Using mass conservation,

\begin{equation}
\frac{\partial \delta}{\partial t}
  = -v \frac{\partial \rho_0}{\partial z}
   - \rho_0 \frac{\partial v}{\partial z},
\end{equation}

\noindent
and assuming that $\delta, v \sim \exp(i \omega t)$, we can
eliminate $\delta$ to get,

\begin{equation}\label{v_pert_eq}
v'' - v' \left[\frac{g +(c^2)'}{c^2}\right]
  + v \left[ \frac{\rho_0''}{\rho_0} 
    - \left(\frac{g +(c^2)'}{c^2}\right)^2
    +\frac{\omega^2}{c^2} \right] = 0,
\end{equation}

\noindent
where $f' = df/dz$.
(Notice that this equation reduces to a sound equation
for very large frequencies.)
Equation \ref{v_pert_eq} can be solved numerically
for the initial density and pressure conditions of the simulation,
for a given frequency $\omega$.
We explored a range of values for $\omega$ and looked for solutions
to equation \ref{v_pert_eq} that are consistent with the zero
vertical velocity boundary conditions we imposed in the simulation.
The lowest frequency consistent with the boundary
conditions corresponds
to a period of about $50 \Myr$ for the vertical gravity and
density structure at $r=8 \kpc$.
As with the analysis from Figure \ref{mass_flux_fourier},
that period decreases as we move inward
($45 \Myr$ at $r=6 \kpc$), and increases outward
($70 \Myr$ at $r=10 \kpc$).
We did not find a period corresponding to the other
$\sim 230 \Myr$ peak, which could involve radial motions.


\begin{figure}[b]
\plotone{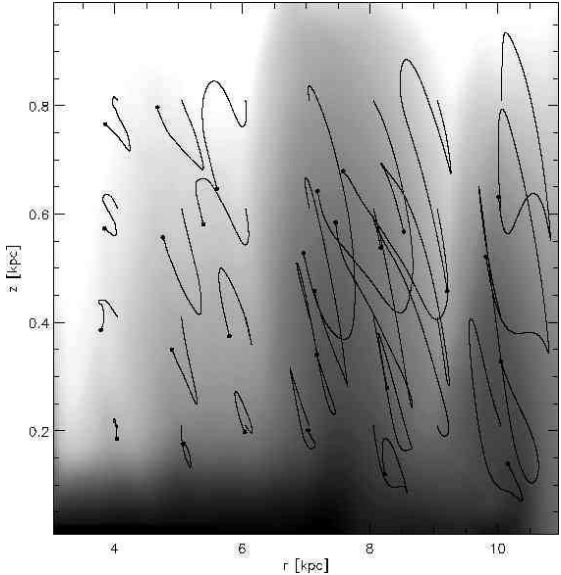}
\caption{
Paths of gas particles as they move from
$\phi=\pi/2$ to $\phi=0$ in the four-arm case,
ending at the position with the dot.
The grayscale is the logarithm of the density at those angles.
The main difference between the looping behavior at low $z$
and the prostate ``s'' shapes higher up is that, at minimum height,
the low $z$ material is moving outward, while the high $z$ material
is moving inward.
In all cases, the general trend is outward when falling
(approaching an arm) and inward when rising (within an arm).
}
\label{paths}
\end{figure}

\section{Circulation.}\label{section_circulation}

In this section, we look for global trends in the gas motions
to explore circulation and radial motions of the material.
In order to do so, we averaged all the velocities and densities of the
simulation for the period of $600 - 1000 \Myr$.
Then, we integrated the resulting mean velocity field in
order to get the trajectory of an imaginary test particle.
The result for the four-arm case
is presented in Figure \ref{paths}.
In this, the grayscale is the time averaged density at $\phi=0$.
The lines show the trajectory in $r$ and $z$ that a test particle
follows in the mean velocity field in going from $\phi=0$ to
$\pi/2$ (one arm encounter time, which differs with radius),
ending at the position with the larger dot.

The first thing to notice in this diagram is that, as the gas
goes between arm and interarm regions, it moves a typical
distance of about $1 \kpc$ in $r$, and a varying amount in
$z$; sometimes, the vertical displacements can be as large
as $0.5 \kpc$.
Also, the trajectories have a slanted appearance.
This is consistent with what has been mentioned before:
at the spiral shock, the gas shoots up and then moves radially
inward along the arm.
After leaving the arm, the gas falls down and moves
out in radius as it approaches the next arm.
The difference between the roughly looping orbits seen at low $z$
and the prostrate ``s'' shapes at higher $z$ is that material at low $z$
is moving outward at minimum height, whereas at high $z$
it is moving inward.
Reference to Figure \ref{four_arm} shows that, at low $z$,
minimum height occurs at the interarm bounce, where material is
moving outward, while at high $z$, minimum height occurs downstream
of the leading edge of the arm, where material has already
begun to move in.

\begin{figure}[b]
\plotone{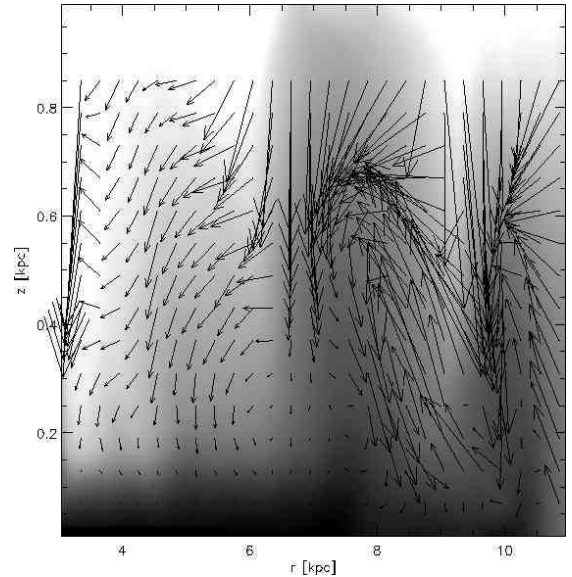}
\caption{
Same as in Figure \ref{paths}, but showing the net
displacement of the particles for a $\Delta \phi=\pi/2$
fraction of their orbit.
The arrows point from to the initial to the final positions.
A counterclockwise cycle around the arms can be seen.
}
\label{arrows}
\end{figure}

In Figure \ref{arrows},
we plotted an arrow from the initial to the final
positions of the test particle.
The interpretation of this diagram is not
straightforward.
Each parcel is followed through one fourth
of a rotation, relative to the pattern.
It typically ends up
at both a different $z$ and $r$, and at a different $r$ it is in a
different phase with respect to the arm, as evidenced by the
shift relative to the density distribution shown.
The pattern
is also incomplete, in that it shows no obvious source of
replenishment for material above $600 \pc$ outward of $6 \kpc$.
What is clear is that
there is a general circulation pattern counterclockwise about each
arm structure, in which
material above $200 \pc$ has large radial and vertical net
excursions.
Higher $z$ material tends to move inward, and lower $z$ outward,
though at lower velocity and higher density.
In the upper regions of the inner edge of each
arm, there is some indication of inward motion that transits
from one arm to the next, but we are reluctant to conclude,
from modeling by this Eulerian code,
that there is thorough mixing of material across the disk.

The fact that the net circulation pattern has several cells,
one associated with each arm, shows that it is
not a simple consequence of
our closed inner and outer boundaries, which might have led to
a single cell of circulation.

This Figure is ambiguous about
what happens to material very close to the plane, leading us to
study the average radial velocity of material as a
function of $z$.
We calculated the mean radial velocity of the gas at
different heights by averaging the radial velocity
at a given $z$, weighted by the density at that point:

\begin{equation}
\bar v_r(z) = \frac{\int \rho v_r ~rd\phi~dr}{\int \rho ~rd\phi~dr}.
\end{equation}

\begin{figure}
\plotone{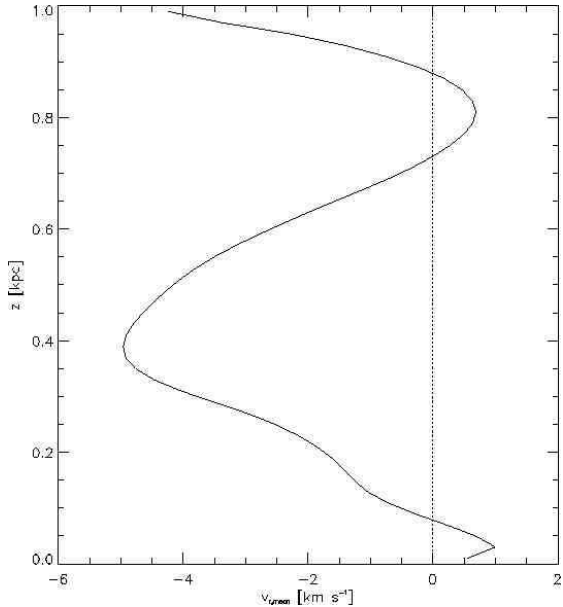}
\caption{
Mean radial velocity for the time-averaged
four-arm case.
The gas moves in above and along the arms, and out in the
interarm region, closer to the plane similar to the magnetic
field pattern in Figure \ref{v.b1}.
The region of slight positive velocity at $z=0.8\kpc$ involves
very little material.
}
\label{radial_flow}
\end{figure}

\noindent
In Figure \ref{radial_flow}, we present this mean
radial velocity for the time averaged four-arm case.
The velocities are inward and significant above $100 \pc$, reaching
$-4 \kms$ at about $400 \pc$.
Near the plane they are outward and low, only about $1 \kms$ is reached.
This is consistent with the observational limits
\citep[and references therein]{por00} and inferences from the
previous Figure.

From this, we conclude, as anticipated, that there is net
outward flow close to the plane to compensate the net inward
flow at higher $z$.
But we are still uncertain whether this
flow occurs in essentially closed cells, separately mixing
material within each arm, as suggested by Figure \ref{arrows}
for most of the material, or whether there is sufficient transit
between arms for substantial global mixing.
Our tentative
conclusion is that the velocities shown in Figure
\ref{radial_flow} are
dominated by closed cell circulation, and therefore represent
a strong limit on the global mixing rate.


\section{Conclusions.}\label{section_conclusions}

We performed MHD simulations of the gaseous galactic disk
having spiral gravitational perturbations with two and four arms.
As found in \citet{mar98} and Paper I, the gas 
motions involve vertical bouncing, plus a combination of
a shock and a hydraulic jump at the gaseous arms.
The comparative timescales for bouncing and interarm passage,
and the realtive phases lead to complications in the details
of the arm structure, as predicted by \citet{mar98} and \citet{wal01}.
This bounce/shock/jump structure
leans upstream above the plane, more noticeably
in the two-arm case, where the bounce point occurs within the arm
at low $z$.
In the four-arm case, the low $z$ bounce point occurs at the
interarm density peak (at $z=0$) and the upward motion from the
bounce leads the shock/jump structure.
In both cases, the gas moves up and over the midplane arm,
inward in radius.
The higher $z$ material then falls down behind the arm,
where it is again moving radially outward.
In the two-arm case, the gas bounces back up and, at high $z$,
generates structures similar to those at the arm, as if
the gas were trying to develop a four-arm structure.
In the four-arm case, if the Sun's position is properly
chosen, the arms seem to follow the arms traced by 
\citet{geo76}.
In both cases, the gaseous arms follow a spiral tighter than
the imposed perturbation.

Since the net radial inflow happens above the arms,
and the gas falls
in the interarm, when it is moving out before encountering the
next arm, the disk averaged radial flow changes sign with height
above the disk, from positive near the midplane, to negative
higher up.
This cycle appears to happen in cells
associated with the arms, and might not represent a global
mixing phenomenon.

Within the spiral arms, the magnetic field adopts a pitch
angle similar to that of the arms, but it develops a negative
pitch in the interarms.
The vertical magnetic field is only important at the position
of the largest vertical flows.
Our model does not include cosmic rays, supernovae or
other energetic events, and therefore, it does not develop
much vertical field, and the total field strength
falls faster with $z$ than observations suggest.
Those shortcomings, on top of our low resolution, did not
allow us to model the random component of the field, which
causes us to 
overestimate the synchrotron polarization fraction.

An examination of the relationship between $B$ and $n$
found that, when the magnetic pressure strongly exceeded the
thermal, $B$ was constant, independent of $n$.
Conversely, when thermal pressure is significant, $B$ tends
to increase with $n$.

The disk atmosphere is close to hydrostatic equilibrium,
when the dynamical terms are taken into account.
It is noticeable that the downward vertical ram pressure of the gas,
when averaged in azimuth, plays an important role
in the hydrostatics, incrementing the effective weight of the gas.
Magnetic tension plays only a small part in the vertical support,
an effect of the model's weak vertical fields, that may not
represent the true situation.

The periodicity of the structures found in this model can
be estimated by Fourier analyzing the vertical mass flux.
We found two outstanding periods, $T \sim 60$ and $\sim 230 
\Myr$, the smaller of which can be explained in terms of the
lowest normal mode of vertical oscillation.

Modelers have the huge advantage of knowing exactly where material
being studied is,
and its velocity.
Observers do not have that luxury.
We will make that connection for our model in 
\citet{gom04},
in which we generate synthetic observations, and then explore
them from the advantageous point of knowing the details
of the underlying distributions.


\acknowledgements

We thank R. Benjamin, E. Zweibel, M. Martos and B. Pichardo and the
anonymous referee for useful comments and suggestions, to the NASA
Astrophysics Theory Program for financial support under the grant
NAG 5-12128, and to M\'exico's Consejo Nacional de Ciencia y
Tecnolog\'{\i}a for support to G. C. G.


{}

\end{document}